\documentclass[prb,aps,11pt,reprint,showkeys,superscriptaddress,longbibliography]{revtex4-1}

\usepackage{amsmath}
\usepackage{amssymb}
\usepackage{graphicx}

\begin{document}
	
	\title{Aharonov-Bohm interference as a probe of Majorana fermions}
	
	\author{T.~C.~Bartolo}
	\email{tommy.bartolo@gmail.com}
	\affiliation{Chemical and Quantum Physics, School of Science, RMIT University, Melbourne, Australia}
	
	\author{J.~S.~Smith}
	\affiliation{Chemical and Quantum Physics, School of Science, RMIT University, Melbourne, Australia}
	
	\author{B.~Muralidharan}
	\affiliation{Department of Electrical Engineering, Indian Institute of Technology Bombay, Powai, Mumbai 
	400076, India}
	
	\author{C.~M\"uller}
	\affiliation{IBM Quantum, IBM Research - Zurich, 8803 R\"uschlikon, Switzerland}
	\affiliation{ARC Centre for Engineered Quantum System, School of Mathematics and Physics, University of 
	Queensland, Brisbane, QLD 4072, Australia}
	
	\author{T.~M.~Stace}
	\affiliation{ARC Centre for Engineered Quantum System, School of Mathematics and Physics, University of 
	Queensland, Brisbane, QLD 4072, Australia}
	
	\author{J.~H.~Cole}
	\email{jared.cole@rmit.edu.au}
	\affiliation{Chemical and Quantum Physics, School of Science, RMIT University, Melbourne, Australia}
	
	\begin{abstract}
		
		Majorana fermions act as their own antiparticle, and they have long been thought to be confined to the realm of pure theory.
		However, interest in them has recently resurfaced, as it was realized through the work of Kitaev that some experimentally accessible condensed matter systems can host these exotic excitations as bound states on the boundaries of 1D chains, and that their topological and non-abelian nature holds promise for quantum computation.
		Unambiguously detecting the experimental signatures of Majorana bound states has turned out to be challenging, as many other phenomena lead to similar experimental behaviour.
		Here, we computationally study a ring comprised of two Kitaev model chains with tunnel coupling between them, where an applied magnetic field allows for Aharonov-Bohm interference in transport through the resulting ring structure. 
		We use a  non-equilibrium Green's function technique to analyse the transport properties of the ring in both the presence and absence of Majorana zero modes.
		Further, we show that these results are robust against weak disorder in the presence of an applied magnetic field.
		This computational model suggests another signature for the presence of these topologically protected bound states can be found in the magnetic field dependence of devices with loop geometries.
		
	\end{abstract}
	
	\maketitle
	
	\section{Introduction}
	
	Majorana fermions were postulated as fermionic excitations that act as their own antiparticle in 1937~\cite{Majorana1937}, but so far they have not been experimentally shown to exist in nature. 
	More recently it was shown theoretically that 1D systems exhibiting $p$-wave superconductivity within 
	certain parameters host Majorana zero modes (MZMs) at their boundaries~\cite{Kitaev2001}.
	It was soon realized that the topological nature of these bound states make them ideal candidates for 
	quantum computation, as it helps to protect them from some types of decoherence~\cite{Alicea2011,Lian2018}.
	Destroying information encoded in this way requires a global perturbation that is strong enough to break the 
	topologically non-trivial phase of the system~\cite{Ivanov2001,Volovik1999}.
	Further, their non-abelian character~\cite{Kitaev2001,Read2000} allows one to manipulate pairs of MZMs 
	through braided exchange of their relative positions, showing a way towards topologically protected 
	quantum computation~\cite{VanHeck2012,Kitaev2001,Alicea2011,Kauffman2016,Kitaev2003}.
	One possible experimental system hosting MZMs are one-dimensional (1D) semiconductor wires with strong 
	spin-orbit coupling and proximity induced superconductivity~\cite{Kitaev2001,Mourik2012}.
	Several recent works have reported experimental evidence for the existence of MZMs in such condensed 
	matter systems~\cite{Zhang2018,Mourik2012,Law2009,Das2012,Deng2012,Rokhinson2012,Deng2016}. 
	However there exist a number of confounding effects that have signatures similar to 
	MZMs~\cite{Liu2018,Liu2017,Deng2016,Moore2018a,Den2018,Chiu2019,Liu2018,Moore2018b} and which 
	make the unambiguous detection of Majorana fermions an ongoing challenge.
	
\begin{figure}[t!]
		\centering
		\includegraphics[width=\columnwidth]{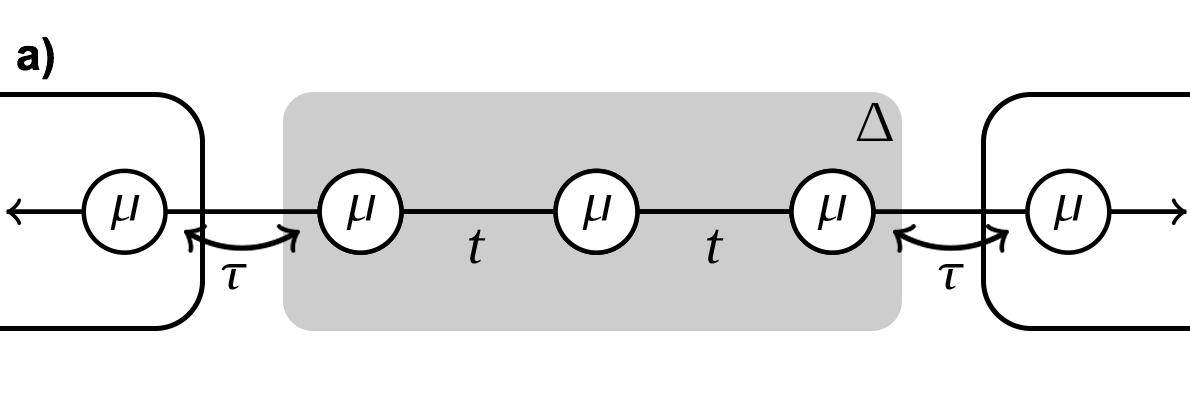}
		\includegraphics[width=\columnwidth]{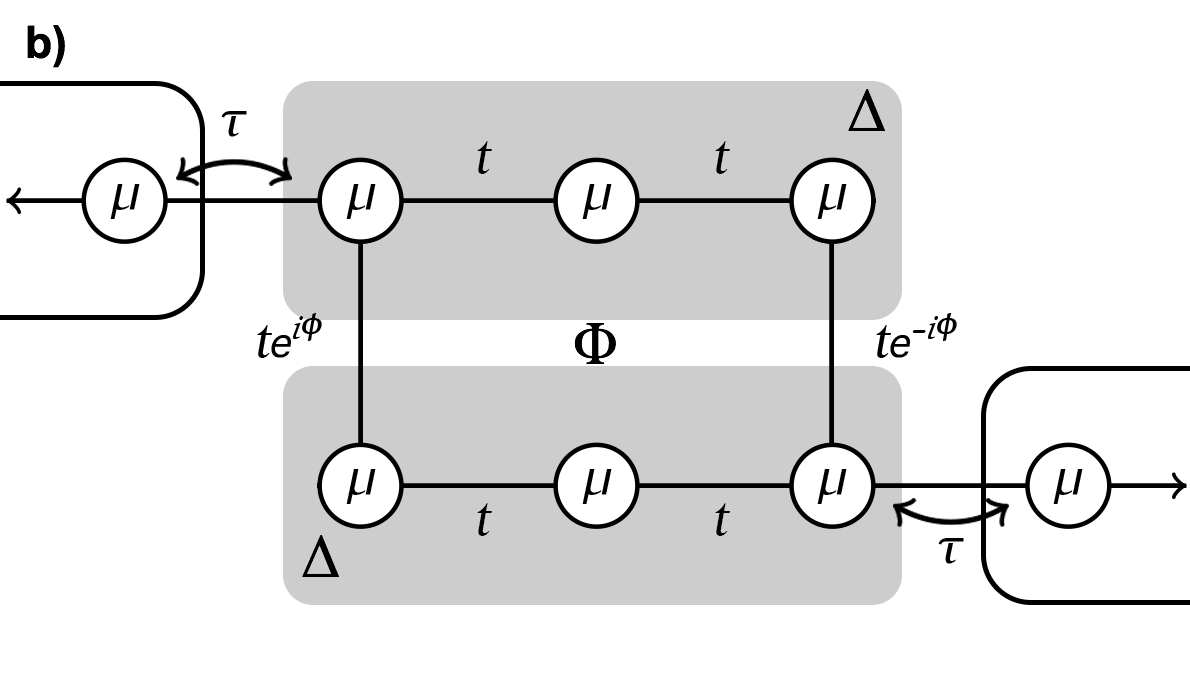}
		\caption{A diagram of a 1D Kitaev nanowire (a) and Aharonov-Bohm ring (b). These consists of a tight 
		binding chain with onsite potential $\mu$, hopping strength $t$, and a $p$-wave pairing amplitude of 
		$\Delta$. Normal leads are coupled to the nanowire with a hopping amplitude of $\tau$.}
		\label{fig:illustrations}
	\end{figure}

	The standard theory model that allows for MZMs is the Kitaev model nanowire~\cite{Kitaev2001}.
	It consists of a 1D tight binding chain with proximity induced $p$-wave superconductivity, as illustrated in 
	Fig.~\ref{fig:illustrations}a.
	Several different materials have been considered to realise such wires~\cite{Kitaev2001,Zhang2018,Mourik2012,Law2009,Das2012,Deng2012,Rokhinson2012,Deng2016}. Although the system parameters vary between the different materials, for all these systems the key physics can be approximated by the effective Kitaev nanowire Hamiltonian.

	The parameter regimes of the Kitaev nanowire have been extensively 
	studied~\cite{Kitaev2001,Chen2014,alicea2012,Lee2014}, as have its transport 
	properties~\cite{Doornenbal2015,Weithofer2014,Danon2020}.
	By coupling two Kitaev nanowires through non-superconducting links at their ends, we form an 
	Aharonov-Bohm (AB) ring, as illustrated in Fig.~\ref{fig:illustrations}b.
	
	Different ring geometries have previously been investigated using interferometry where it was suggested the 
	periodicity of the conductance as a function of magnetic field might be used as a way to identify Majorana 
	bound states~\cite{VanHeck2011,Zeng2017,Akhmerov2011,Whiticar2020}.
	Other studies employed either scattering matrix theory or Green's function techniques to study a normal AB 
	ring containing a single nanowire hosting MZMs~\cite{Tripathi2016,Shang2014,Benjamen2010,Hansen2016,Jiang2015}.
	For example, the transport properties of such an AB ring were found to be sensitive to the difference between 
	MZMs and Andreev bound states~\cite{Tripathi2016,Haug2019a,Haug2019b,Haug2019c,Qin2018,Aksenov2019,Aksenov2020}.
	Studies of two-dimensional systems of finite extent have investigated interference effects due to chiral 
	Majorana fermion edge states at the normal-superconductor boundary~\cite{Park2014,Fu2009,Law2009}.
	
	In this paper we study the interplay between the AB effect and MZMs, and expand on these previous works by 
	analysing the transport characteristics of an AB ring formed by two coupled Kitaev nanowires. 
	We employ the non-equilibrium Green's function (NEGF) formalism and explore the relationship between 
	model parameters and transport characteristics for a finite size ring geometry.
	Previous studies have investigated the case of MZMs in a finite nanowire~\cite{Zvyagin2015,Leumer2020}.
	Such models are particularly useful as all experimentally realizable devices are finite, which limits what can be 
	understood simply from the bulk properties of the materials hosting MZMs.
	The relation between AB interference and Majorana bound states in such a loop geometry has previously 
	been studied using a scattering matrix approach with the wide-band 
	approximation~\cite{Ya-Jing_Zhang2018}. 
	However energy resolved transport characteristics of such a circuit provide important clues on how such 
	effects can be probed experimentally and will help to further understand the experimental signatures of 
	these topologically non-trivial bound states. Here we show that mapping the energy resolved transmission 
	through an AB ring comprised of two MZMs displays the expected resonance at zero energy. However, 
	mapping this resonance as a function of magnetic field, on-site potential or superconducting order 
	parameter results in characteristic responses, suggesting the possibility of unambiguously distinguishing 
	MZM from trivial bound states.
	
	The paper proceeds as follows: Section~\ref{sec:kitaev_nanowire} introduces the standard toy model for the 
	Kitaev nanowire.
	The NEGF formalism is then outlined in Section~\ref{sec:wire_negf}. 
	Section~\ref{sec:AB_ring} expands on the Kitaev nanowire by introducing an AB ring comprised of two 
	coupled Kitaev nanowires. 
	Finally, Section~\ref{sec:AB_ring_magfield} examines the transport properties of this ring in the presence of 
	an applied magnetic field.
	We conclude in Section~\ref{sec:conclusion}.
	
	\begin{table}[b!]
		\caption{The different parameter regimes for the Kitaev nanowire. These regimes are obtained from the 
		Hamiltonian shown in Eq.~(\ref{eq:KitaevToyH}).}
		\begin{tabular}{l c}
			\hline \hline
			Regime & Model parameters \\ \hline
			Topologically trivial & $\mu/t > 2$ \\
			Topologically non-trivial & $\mu/t < 2$, $\Delta/t \neq 0$ \\
			Superconducting & $\Delta/t >> 1$ \\
			Normal & $\Delta/t = 0$ \\
			\hline \hline
		\end{tabular}
		\label{tab:regimes}
	\end{table}
	
	\begin{figure}[t!]
		\includegraphics[width=\columnwidth]{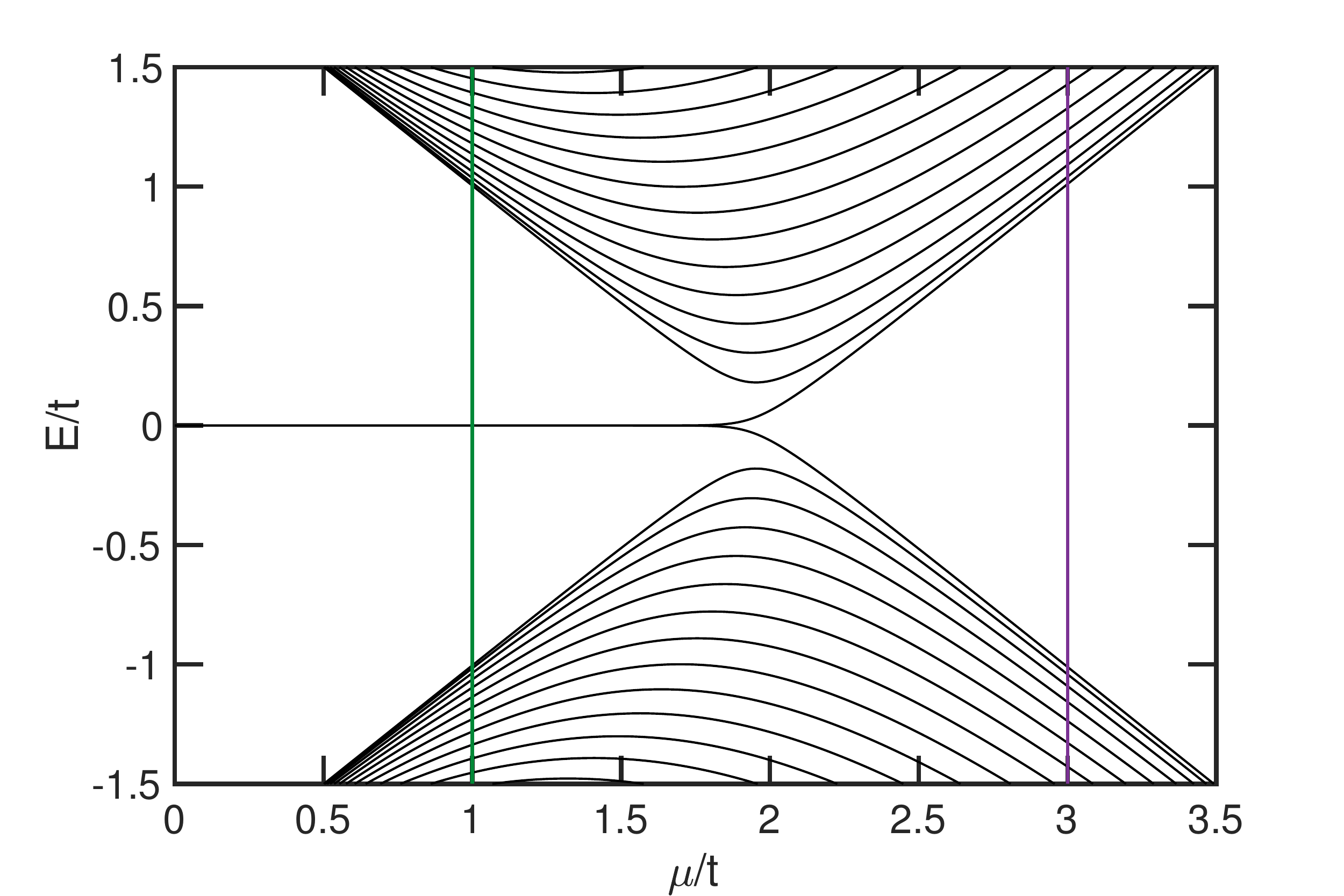}
		\caption{Eigenenergies for a 1D Kitaev nanowire as a function of $\mu / t$ with $\Delta/t = 1.0$. Vertical 
		lines indicate values of $\mu/t$ used to compute the transmission probabilities in 
		Fig.~\ref{fig:1dchain_spectrum_delta1_mu2o5}.}
		\label{fig:1dchain_spectrum_delta1}
	\end{figure}
	
	\begin{figure}[t!]
		\includegraphics[width=\columnwidth]{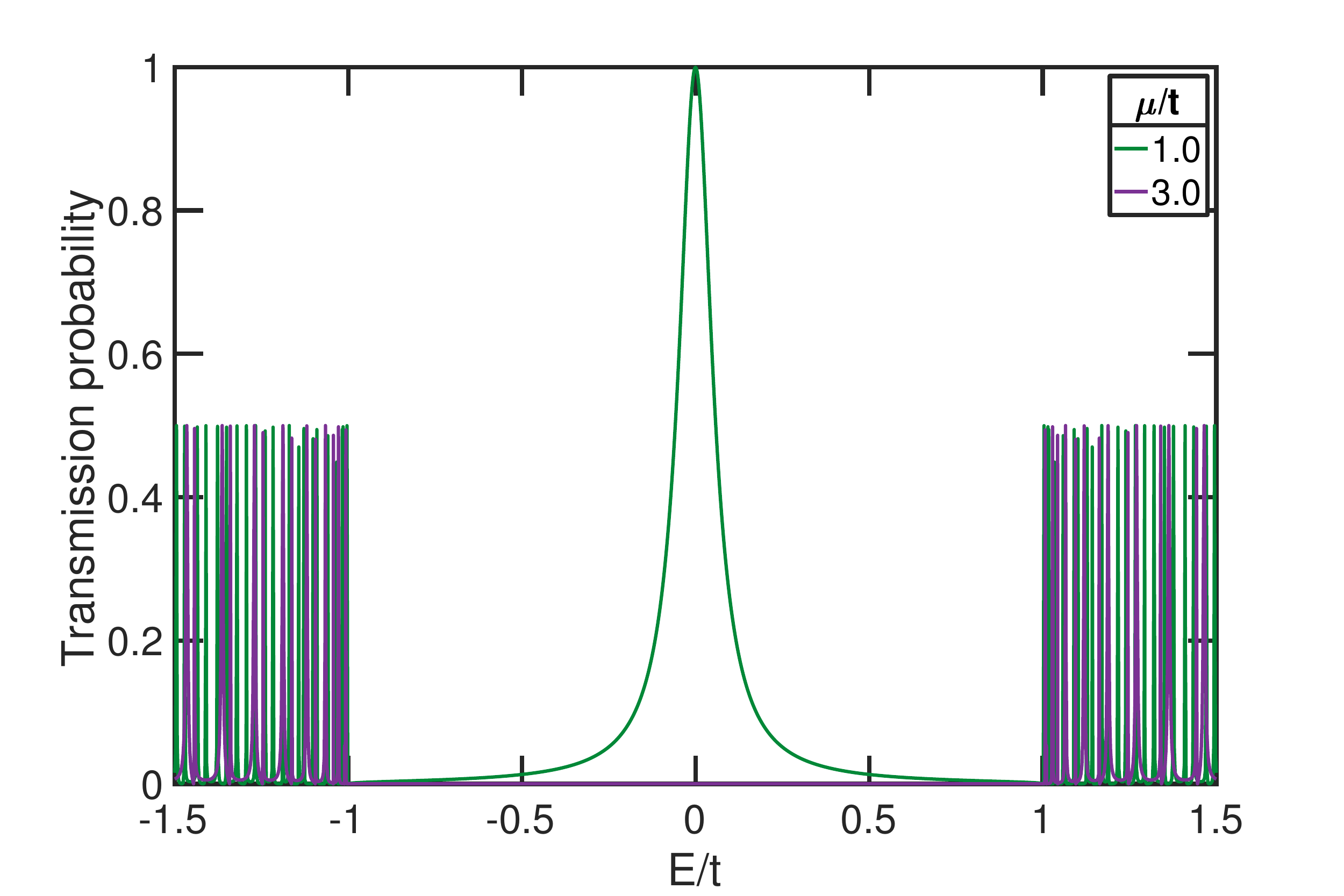}
		\caption{Transmission probability as a function of energy for the 1D Kitaev nanowire for two values of 
		$\mu/t$, highlighted in Fig.~\ref{fig:1dchain_spectrum_delta1} as vertical lines. The nanowire has a 
		$p$-wave pairing amplitude of $\Delta/t = 1.0$ and a hopping strength between the leads and the device 
		of $\tau/t = 0.3$.}
		\label{fig:1dchain_spectrum_delta1_mu2o5}
	\end{figure}
	
	\section{The Kitaev nanowire} \label{sec:kitaev_nanowire}
	
	We begin by following Kitaev in defining a model Hamiltonian ($\hat{H}_{\rm{KC}}$) for a system admitting the existence of MZMs as

			\begin{align}
			\hat{H}_\mathrm{NW} &= \sum^{\mathrm{N}}_{j=1} \bigg[-t \left(c_j^\dagger c_{j+1}\right) 
			- \mu\left(c_j^\dagger c_j - \frac{1}{2}\right) + \mathrm{h.c.}\ \bigg] \nonumber \\
			\hat{H}_\mathrm{KC} &=  \hat{H}_{\mathrm{NW}} + \sum^{\mathrm{N}}_{j=1} \bigg[\Delta e^{i\theta} c_j c_{j+1} + \Delta e^{-i\theta} c_{j+1}^\dagger c_{j}^\dagger \bigg] \label{eq:KitaevToyH}
			\end{align}
			
	where $t$ is the nearest-neighbour hopping strength, $\mu$ is the on-site potential, $\Delta$ is the 
	$p$-wave pairing amplitude, $\theta$ is the superconducting phase and N is the number of sites in the chain.
	Without loss of generality we set $\theta = 2\pi n$ where $n \in \mathbb{Z}$, resulting in $e^{i\theta} = 1$.
	$c_j$ is the fermionic annihilation operator acting on the ${j\mathrm{th}}$ site.
	We write a particle-hole symmetric (PHS) form of the Hamiltonian through the Bogoliubov-deGennes (BdG) 
	Hamiltonian:
	\begin{align}
	H_{\mathrm{BdG}} =
	\begin{bmatrix}
	H_{\mathrm{NW}} & \Delta \\
	-\Delta^* & -{H_{\mathrm{NW}}^{*}}
	\end{bmatrix}
	\end{align}
	$H_{\mathrm{NW}}$ is the Hamiltonian for a normal wire and $\Delta$ is the $p$-wave pairing amplitude 
	between particles and holes.
	For more details see App.~\ref{appendix:ParticleHoleSymmetry}.
	Following Chen \textit{et al.}\cite{Chen2014} we model the Kitaev nanowire as a 1D tight binding chain that consists of 50 sites with the on-site potential $\mu$ and $p$-wave pairing amplitude $\Delta$ as adjustable parameters.
	The essential physics studied here remains unchanged with the length of the chain, so long as the chain is long enough such that the groundstate wavefunctions do not overlap, as is the requirement for topologically non-trivial MZMs.
	To calculate the electrical response of the 1D Kitaev nanowire we couple it to normal leads on either side and 
	apply an NEGF technique, as is described in Sec.~\ref{sec:wire_negf}.
	
	The Kitaev Hamiltonian has three different regimes which are of interest to us here, these are the 
	topologically trivial, the topologically non-trivial, and the superconducting regime.
	The different model parameters for each of these regimes are summarized in Tab.~\ref{tab:regimes}, where 
	we also list parameters for the normal (non-superconducting) regime.
	Fig.~\ref{fig:1dchain_spectrum_delta1} shows how the eigenspectrum of this Hamiltonian depends on the 
	value of $\mu/t$.
	In this figure we see the separation of the topologically trivial and non-trivial regimes at approximately 
	$\mu/t = 2$.
	When $\mu/t < 2$ the system is topologically non-trivial and we have a degenerate ground state at $E/t = 0$.
	In the topologically trivial regime, when $\mu/t > 2$, the ground state of the system is no longer degenerate, 
	nor does it appear at $E/t = 0$.
	
	\subsection{Non-equilibrium Green's function formalism}
	\label{sec:wire_negf}
	
	To calculate the electrical response of a Kitaev nanowire we apply the NEGF formalism~\cite{Datta}, which 
	allows us to compute transmission probabilities as a function of energy.
	We are also thereby able to introduce open boundary conditions and model transport through the device with 
	changing magnetic field.
	The retarded Green's function for the device $G^{\mathrm{R}}$, from which the transport properties of the 
	system can be calculated, is written as
	\begin{align}
	G^{\mathrm{R}}(E) = \left[\left(E + i0^{+}\right)\mathbb{I} - H_{\mathrm{BdG}} - \Sigma_{\mathrm{L}}(E) - 
	\Sigma_{\mathrm{R}}(E)\right]^{-1}
	\end{align}
	where $E$ is energy and $0^{+}$ is a small positive number.
	For brevity we suppress the energy dependence in subsequent Green's functions, transmission functions, and 
	other operators.
	The Hamiltonian describing the device is modified by two self-energy terms $\Sigma_{L}$ and $\Sigma_{R}$ 
	to include the left and right leads respectively.
	These self-energy terms are calculated from the hopping parameter and onsite potential for the leads and are used to model the open boundaries.
	The nearest-neighbour hopping strength between the leads and the device is given by $\tau/t = 0.3$.
	These weak links ($\tau < t$) at the interfaces, together with the self-energy terms, modify the system's density of states such that resonant 
	tunneling dominates transport through the device.
	The Green's function is used along with the Caroli formula (Eq.~(\ref{eq:Caroli})) to calculate transmission 
	probabilities for particles/holes moving through the device.
	When applying the NEGF formalism to a conventional charge transport problem, the transmission probability 
	$T$ is given by
	\begin{align}
	T = \mathrm{Tr}\left(\Gamma_{\mathrm{L}} G^{\mathrm{R}} \Gamma_{\mathrm{R}} G^{\mathrm{A}}\right) 
	\label{eq:Caroli}
	\end{align}
	with $G^{\mathrm{A}} = (G^{\mathrm{R}})^\dagger$ and broadening matrices $\Gamma_{\alpha}$ given by
	\begin{align}
	\Gamma_{\alpha} = i\left(\Sigma_{\alpha}-\Sigma_{\alpha}^{\dagger}\right)
	\end{align}
	with $\alpha$ an index describing the left (L) and right (R) contacts.
	
	\begin{figure}[t!]
		\includegraphics[width=0.9\columnwidth]{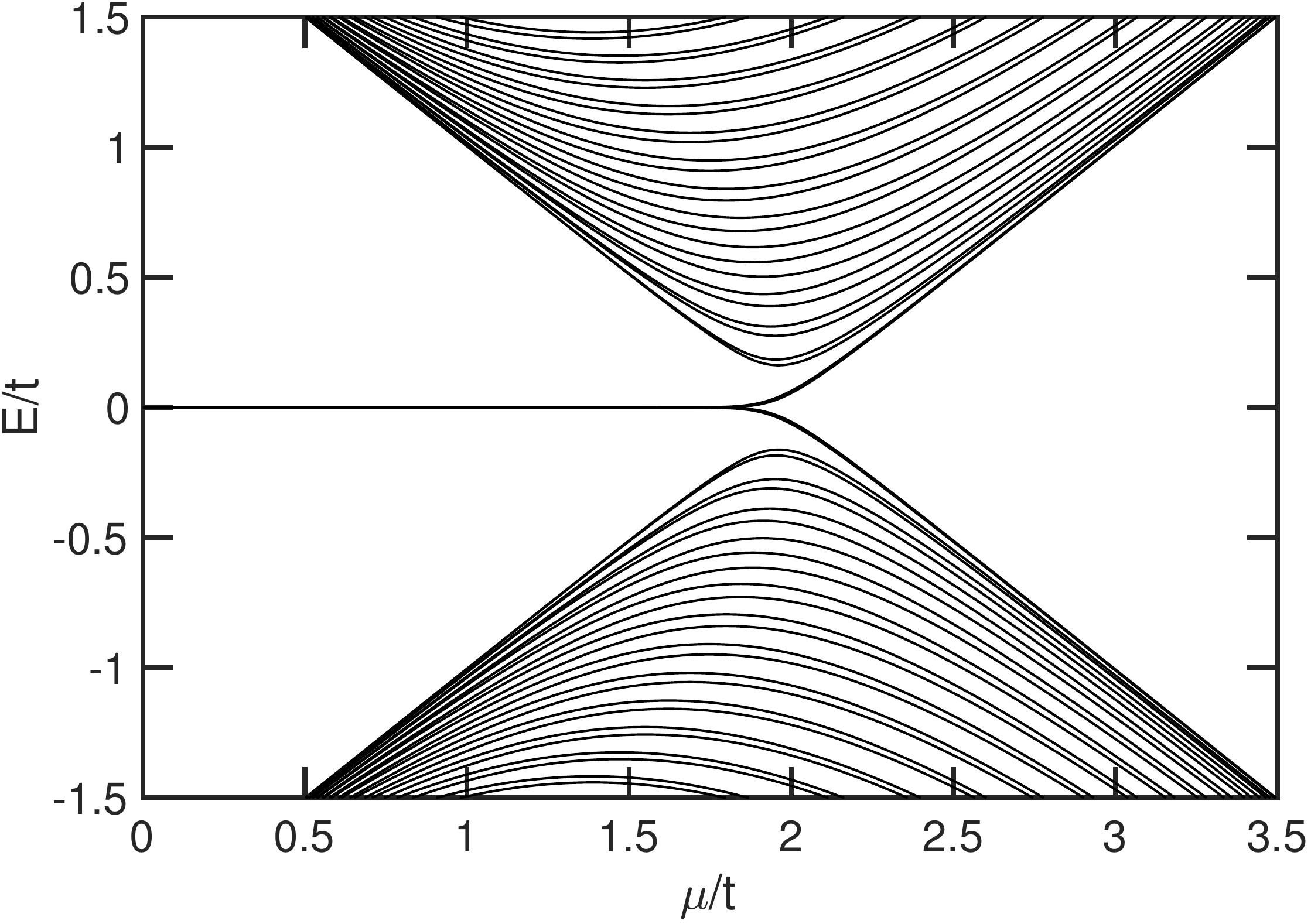}
		\caption{Eigenenergies for an AB ring (without leads attached) as a function of $\mu/t$ with $\Delta/t = 
		1.0$. Here the magnetic flux through the ring is $\Phi/\Phi_{0} = 0$.}
		\label{fig:eigenspectra_halfpi_nontrivial}
	\end{figure}
	
	\begin{figure}[t!]
		\includegraphics[width=\columnwidth]{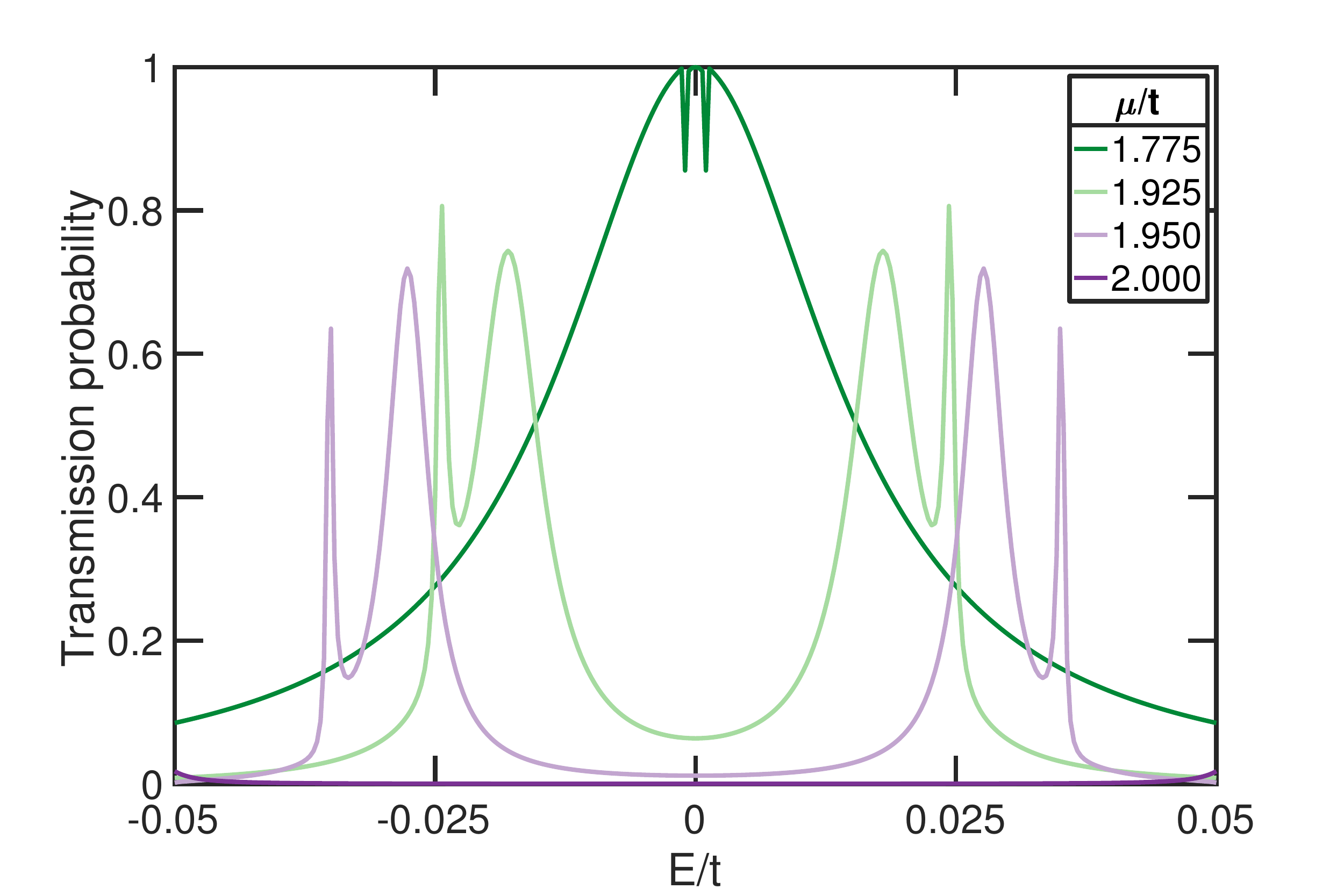}
		\caption{Transmission probability of an AB ring at energies about zero for several different values of the 
		on-site potential $\mu$. The AB ring has a $p$-wave pairing amplitude $\Delta/t = 1.0$ and normal leads 
		are attached with a coupling strength of $\tau/t = 0.3$. The external applied magnetic field strength here 
		is $\Phi/\Phi_{0} = 0$.}
		\label{fig:ZBA_broadening_ring}
	\end{figure}

	The conductance $g$ can be calculated via\cite{Lundstrom2000}
	\begin{align}
		g = \frac{-2q^2}{h}\int_{-\infty}^{\infty} T(E) \frac{\partial f(E)}{\partial E} dE \label{eq:conductance}
	\end{align}
	where $h$ is Planck's constant, $q$ is the charge of the particle, and $f(E)$ is the equilibrium Fermi-Dirac distribution function. In the limit as temperature goes to zero, the conductance is simply proportional to the transmission function and therefore in everything that follows we only consider $T(E)$.
	In general if a voltage bias is applied to the device then the onsite potential becomes spatially dependent and the integral in Eq.~\ref{eq:conductance} must be taken over a range of energies.
	Although these complications are important in discussing real devices they don't change the underlying physics and so we focus on the zero bias limit.
	For the BdG Hamiltonian we must now account for both electron and hole degrees of freedom. We can 
	compute the total transmission probability as the sum of the transmission probabilities for direct 
	transmission $T_{\mathrm{D}}$, Andreev reflection $T_{\mathrm{A}}$, and crossed Andreev reflection 
	$T_{\mathrm{CA}}$:
	\begin{align}
	T^{\mathrm{e(h)}}= T^{\mathrm{e(h)}}_{\mathrm{D}} + T^{\mathrm{e(h)}}_{\mathrm{A}} + 
	T^{\mathrm{e(h)}}_{\mathrm{CA}}
	\label{eq:T_sum}
	\end{align}
	with
	\begin{align}
	T_{\mathrm{D}}^{\mathrm{e(h)}} &= \mathrm{Tr}\left(\Gamma^{\mathrm{e(h)}}_{\mathrm{L}} 
	G^{\mathrm{R}} \Gamma^{\mathrm{e(h)}}_{\mathrm{R}} G^{\mathrm{A}}\right) \label{eq:T_D} \\
	T_{\mathrm{A}}^{\mathrm{e(h)}} &= \mathrm{Tr}\left(\Gamma^{\mathrm{e(h)}}_{\mathrm{L}} 
	G^{\mathrm{R}} \Gamma^{\mathrm{h(e)}}_{\mathrm{L}} G^{\mathrm{A}}\right) \label{eq:T_A} \\
	T_{\mathrm{CA}}^{\mathrm{e(h)}} &= \mathrm{Tr}\left(\Gamma^{\mathrm{e(h)}}_{\mathrm{L}} 
	G^{\mathrm{R}} \Gamma^{\mathrm{h(e)}}_{\mathrm{R}} G^{\mathrm{A}}\right) \label{eq:T_CA}
	\end{align}
	
	Eq.~(\ref{eq:T_sum}) therefore gives the transmission probability of a particle (hole) through the device at a 
	particular energy (see App.~\ref{appendix:ComponentsOfTransmission} for further details).
	The broadening matrices are now represented as a $2\times2$ matrix due to PHS:
	\begin{align}
	\Gamma_{\alpha} =
	\begin{bmatrix}
	\Gamma_{\alpha}^{(\mathrm{e})} & \Gamma_{\alpha}^{(\mathrm{eh})} \\
	\Gamma_{\alpha}^{(\mathrm{he})} & \Gamma_{\alpha}^{(\mathrm{h})}
	\end{bmatrix}
	\end{align}
	where the superscript represents the type of particles involved in the interaction.
	For example (e) is a electron-electron interaction and (h) is a hole-hole interaction.
	As the system has normal leads we do not consider the off-diagonal terms which correspond to electron-hole 
	interaction in the broadening matrices, i.e. $\Gamma_{\alpha}^{\mathrm{(he)}} = 
	\Gamma_{\alpha}^{\mathrm{(eh)}} = 0$.
	As we are ultimately interested in the total conductance through the device, in what follows we often plot the 
	combined transmission due to electrons and holes, $(T^{\mathrm{e}} + T^{\mathrm{h}})/2$, where the 
	individual electron and hole transmission functions are given by Eq.~\ref{eq:T_sum}.
	
	\subsection{Signatures of Majorana fermions and the zero bias anomaly}
	
	A signature of MZMs in the electrical response of these systems is the existence of a zero-bias anomaly (ZBA) 
	that is topologically protected~\cite{Das2012}.
	This ZBA is a conduction mode at $E/t = 0$ that results from the zero-energy ground state of the MZMs.
	Fig.~\ref{fig:1dchain_spectrum_delta1_mu2o5} shows the transmission probability for the 1D Kitaev nanowire 
	in the topologically trivial and the topologically non-trivial regime.
	The figure shows the transmission probability for a particle through the Kitaev nanowire at $E/t = 0$, with 
	on-site potentials $\mu/t = 1$ ($\mu/t < 2$, topologically non-trivial) and $\mu/t = 3$ ($\mu/t > 2$, 
	topologically trivial).
	The presence of a non-zero probability for transmission of a particle at $E/t = 0$ when $\mu/t = 1$ is 
	characteristic of the ZBA\cite{Mourik2012}.
	Here the width of the zero energy peak in the transmission probability is directly related to the strength of the coupling to the device, $\tau$.
	As $\tau$ increases there is a stronger connection between the device and the leads, which creates a broader resonance peak\cite{Datta}.
	In the topologically trivial regime ($\mu/t = 3$) there is no zero energy peak as all states of the Hamiltonian 
	fall outside of the energy gap, as seen in Fig.~\ref{fig:1dchain_spectrum_delta1_mu2o5}.
	
	\section{The Aharonov-Bohm ring} \label{sec:AB_ring}
	
	We now consider a device comprised of two 1D Kitaev nanowires, as illustrated in Fig.~\ref{fig:illustrations}.
	This system is capable of supporting two pairs of MZMs (one MZM at each of the four interfaces between the 
	device and the leads).
	For all AB ring devices modelled here we use a ring comprised of two Kitaev nanowires with 50 sites,
	
	\begin{align}
		H_\mathrm{AB}=
		\begin{bmatrix}
			H_\mathrm{BdG} &  te^{-i\phi} \Pi \\
			 te^{i\phi} \Pi & H_\mathrm{BdG}
		\end{bmatrix} \label{eq:Hring}
	\end{align}

	where the off-diagonal blocks $\Pi$ are the adjacency matrices defined such that they couple the BdG Hamiltonians at the edges of the wires, as per Fig.~\ref{fig:illustrations}b and $\phi$ is the Peierls phase for an applied magnetic field. 
	
	\begin{figure}[t!]
		\includegraphics[width=\columnwidth]{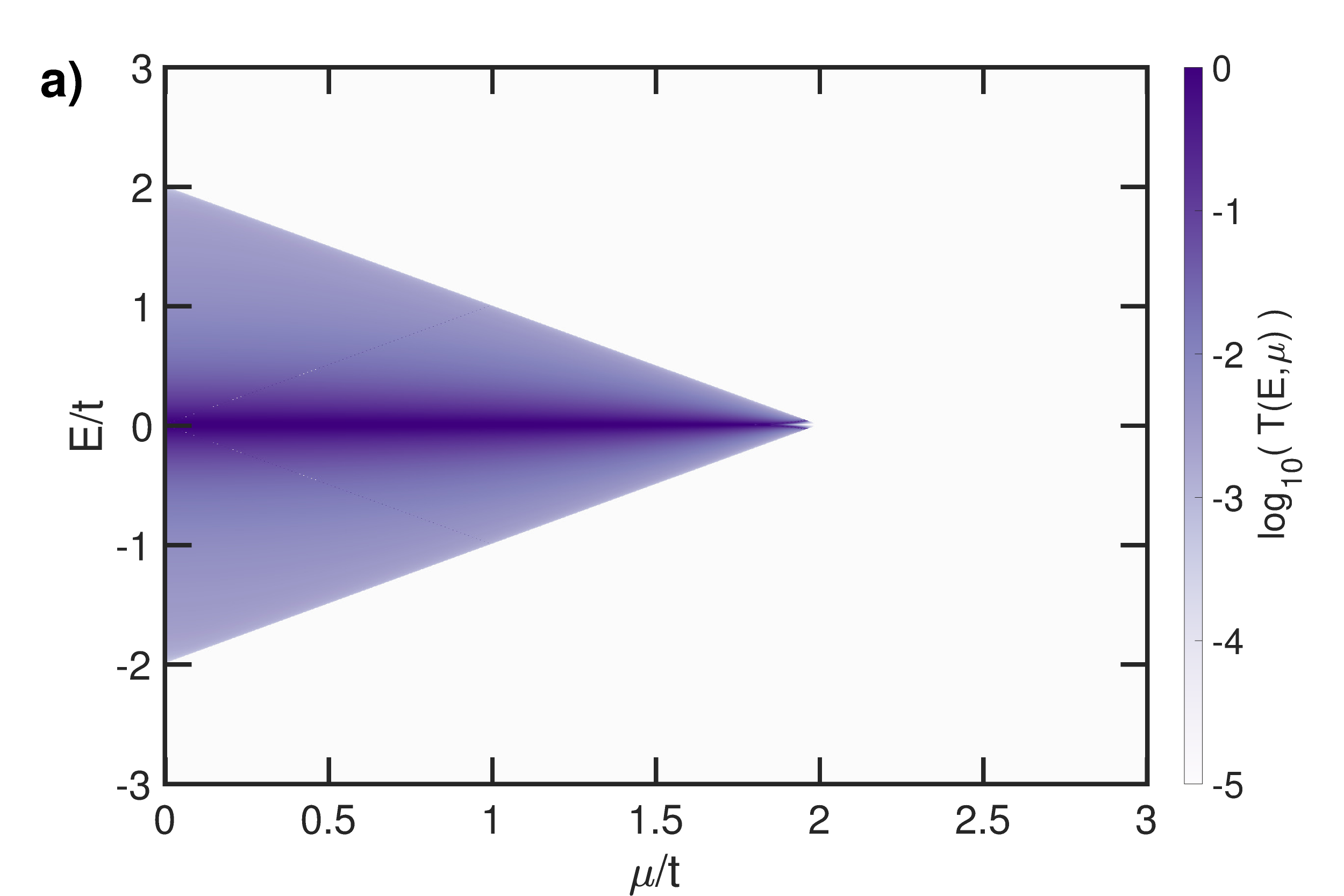}
		\includegraphics[width=\columnwidth]{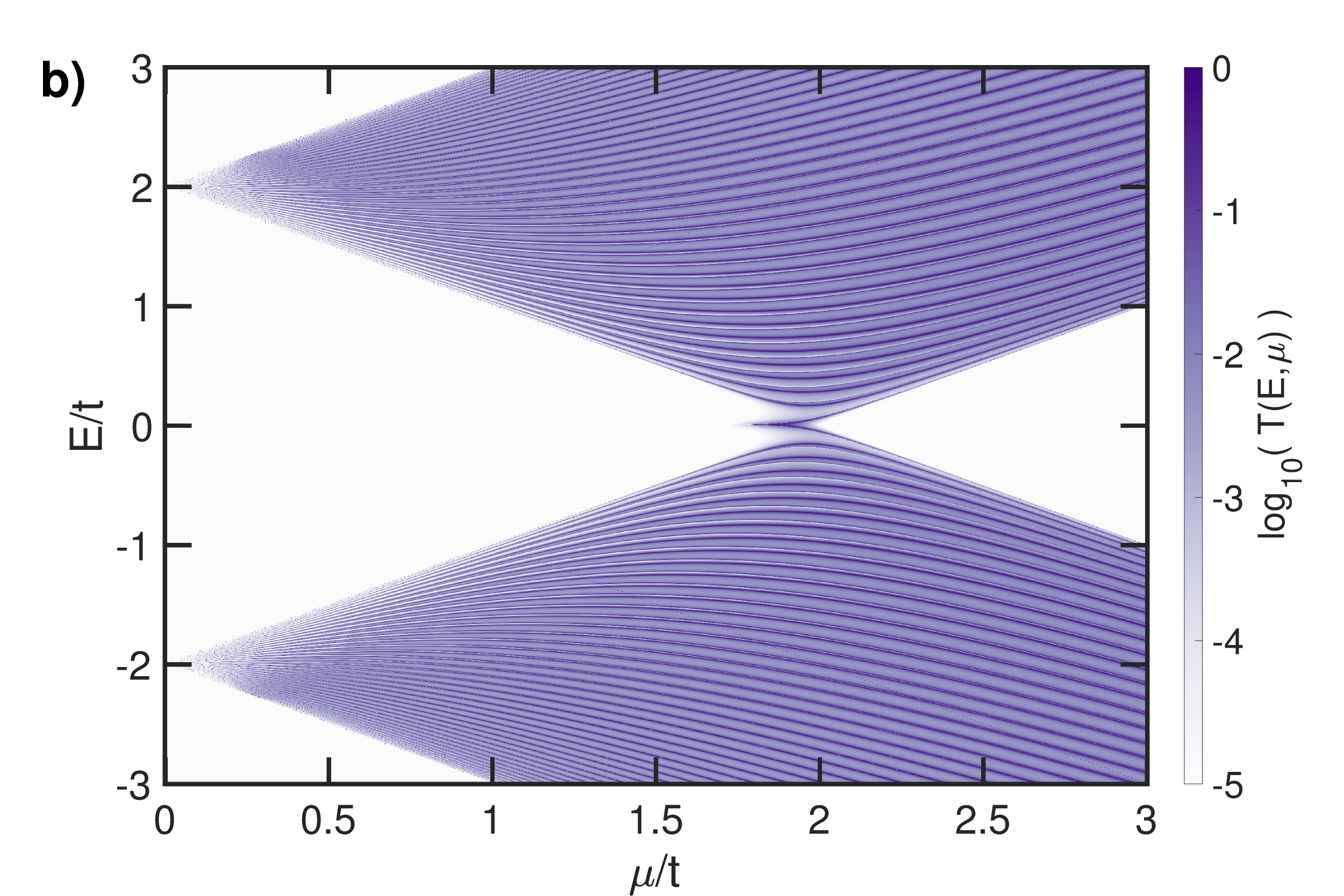}
		\includegraphics[width=\columnwidth]{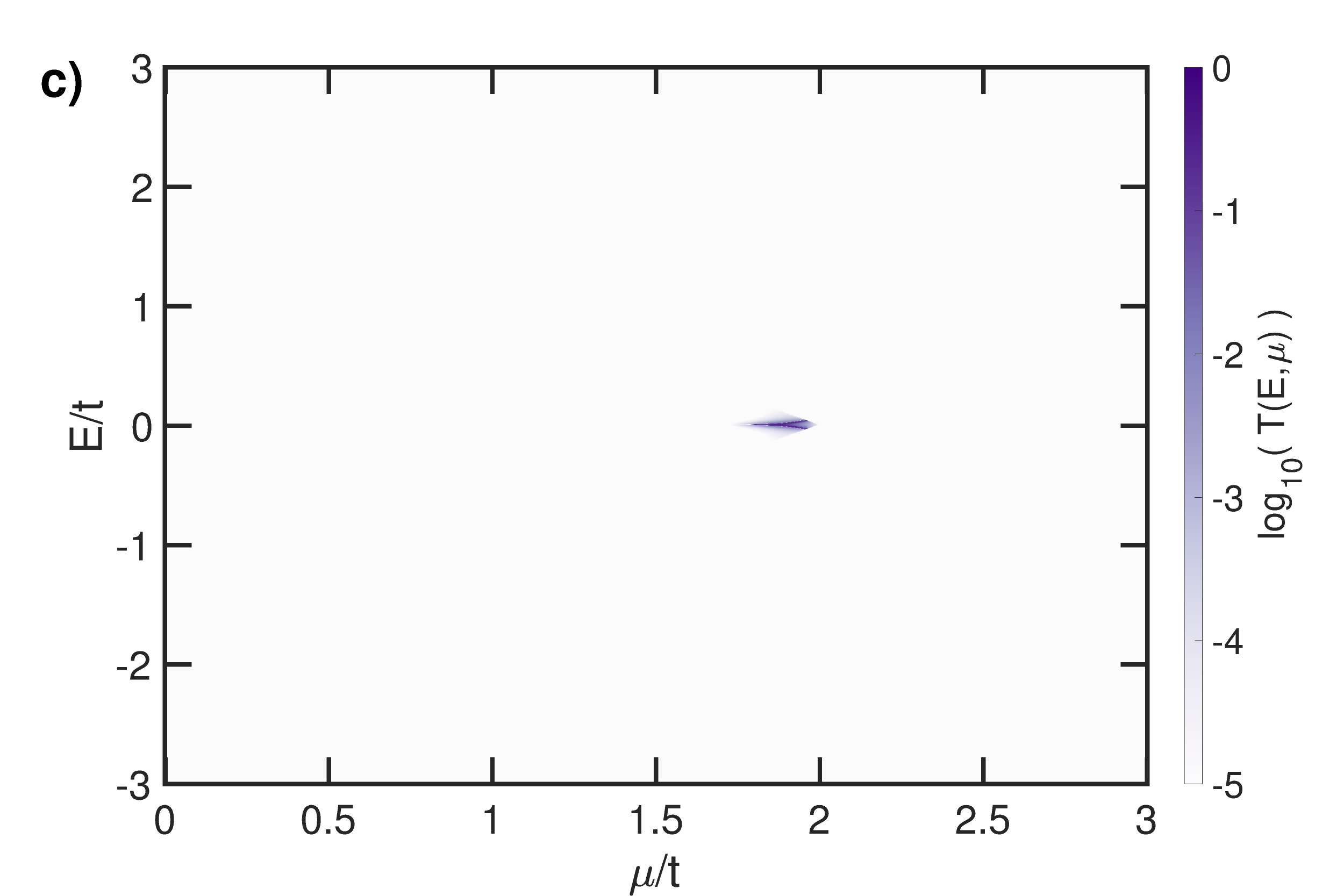}
		\caption{Contributions to the transmission probability from (a) Andreev, (b) direct, and (c) crossed Andreev 
		transmission for an AB ring with $50\times2$ sites as a function of energy $E$ and on-site potential 
		$\mu$. Here the magnetic flux through the ring is $\Phi/\Phi_{0} = 0$. The AB ring has a $p$-wave pairing 
		amplitude $\Delta/t = 1.0$ and normal leads are attached with a coupling strength of $\tau/t = 0.3$. Note 
		the logarithmic color scale used in order to resolve the weaker transmission probabilities in (b) and (c).}
		\label{fig:transmission_breakdown}
	\end{figure}
	
	The topological phase diagram for this AB ring is equivalent to the topological phase diagram of the 1D Kitaev 
	nanowire.
	The presence of a ZBA for the AB ring can been seen in Fig.~\ref{fig:eigenspectra_halfpi_nontrivial} for each 
	of the parameter regimes described in Tab.~\ref{tab:regimes}.
	The two Kitaev nanowires of the ring are connected by a nearest-neighbour hopping strength of $t'$, as 
	shown in Fig.~\ref{fig:illustrations}b.
	In all that follows we set $t'=t_1=t_2=t$.
	The $p$-wave pairing amplitude between the two nanowires is zero.
	This corresponds to the case of normal tunneling between two superconducting nanowires.
	
	We now apply our NEGF method to an AB ring.
	Fig.~\ref{fig:ZBA_broadening_ring} shows the transmission probability for an AB ring at energies close to $E 
	= 0~t$ and values of the on-site potential $\mu/t = \{ 1.775, 1.925, 1.950, 2.000\}$.
	For $\mu/t > 1.775$ we see peaks form due to the topologically trivial quasi-particle states which lie outside the superconducting energy gap.
	As $\mu$ is decreased, we see the formation of a peak in the transmission probability at zero energy.
	The appearance of the ZBA correlates with the system becoming topologically non-trivial.
	At $\mu/t = 1.775$ the zero energy transmission is maximised which suggests that at this value of the 
	on-site potential the system is in the topologically non-trivial regime.
	The value of $\mu$ at which the system becomes topologically non-trivial is affected by the number of sites.
	As the number of sites increases the system approaches the ideal case and the value tends to $\mu/t \approx 2$. 
	For relatively small numbers of sites, the value of $\mu/t$ corresponding to a topological phase transition is reduced due to finite size effects.
	
	By examining each of the individual components of Eq.~(\ref{eq:T_sum}) (given in 
	Eq.(\ref{eq:T_D})-(\ref{eq:T_CA})) we are able to see how each component of  the transmission probability 
	influences the total transmission probability through the device.
	It also allows for insight into how the magnetic field induced interference affects each of the difference types 
	of transmission (direct, Andreev and crossed Andreev).
	Fig.~\ref{fig:transmission_breakdown} shows the transmission probability as a function of energy $E$ and 
	on-site potential $\mu$ for Andreev transmission, direct transmission, and crossed Andreev transmission 
	with a magnetic flux through the ring of $\Phi/\Phi_{0} = 0$.
	We see in Fig.~\ref{fig:transmission_breakdown}a the sub-gap states are due entirely to resonant Andreev 
	transmission, while in  Fig.~\ref{fig:transmission_breakdown}b we see that direct transmission occurs via 
	states above and below the gap.
	Fig.~\ref{fig:transmission_breakdown}c shows that the crossed Andreev transmission only exists close to the 
	point of transition between trivial and non-trivial topological behaviour, at zero energy and 
	$\mu/t\approx2$.
	The crossed Andreev transmission is relatively weak compared to the other contributions.
	
	\section{Magnetic field applied to an Aharonov-Bohm ring} \label{sec:AB_ring_magfield}
	
	We now apply a perpendicular magnetic field to the AB ring by Peierls substitution to observe interference in 
	the transmission probability~\cite{Peierls1933,Maska2001,Hofstadter1976}.
	We choose a gauge such that the Peierls phase $\Phi$ (equivalent to magnetic flux through the ring and 
	given here in units of flux quanta $\Phi_0=h/2e$) drops across only the normal links.
	By applying a magnetic field to the AB ring in the topologically trivial and non-trivial regimes, we see the 
	changes in the transmission probability as a function of magnetic flux due to the presence (or absence) of the 
	MZMs.
	
\begin{figure}[t!]
		\includegraphics[width=\columnwidth]{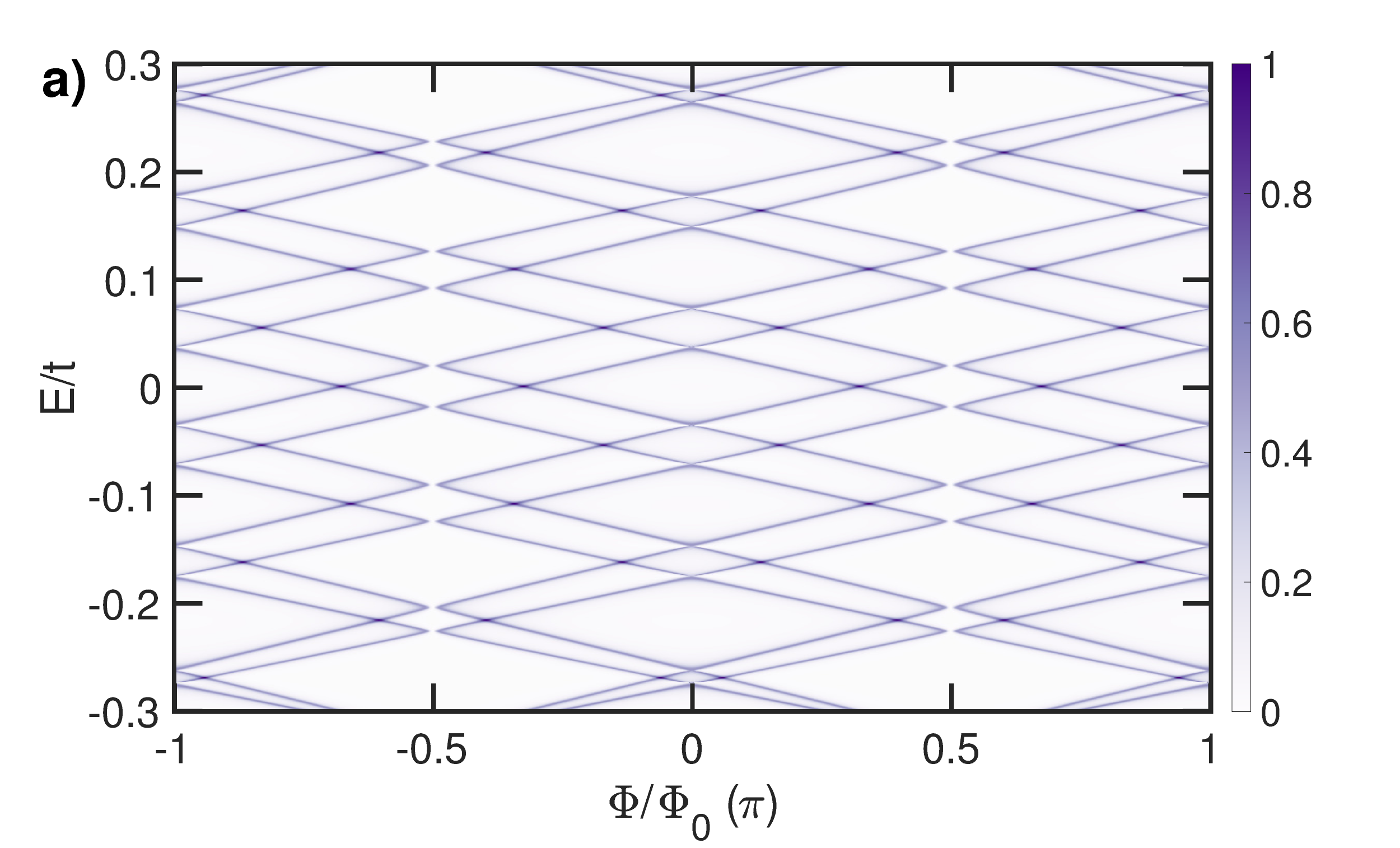}
		\includegraphics[width = \columnwidth]{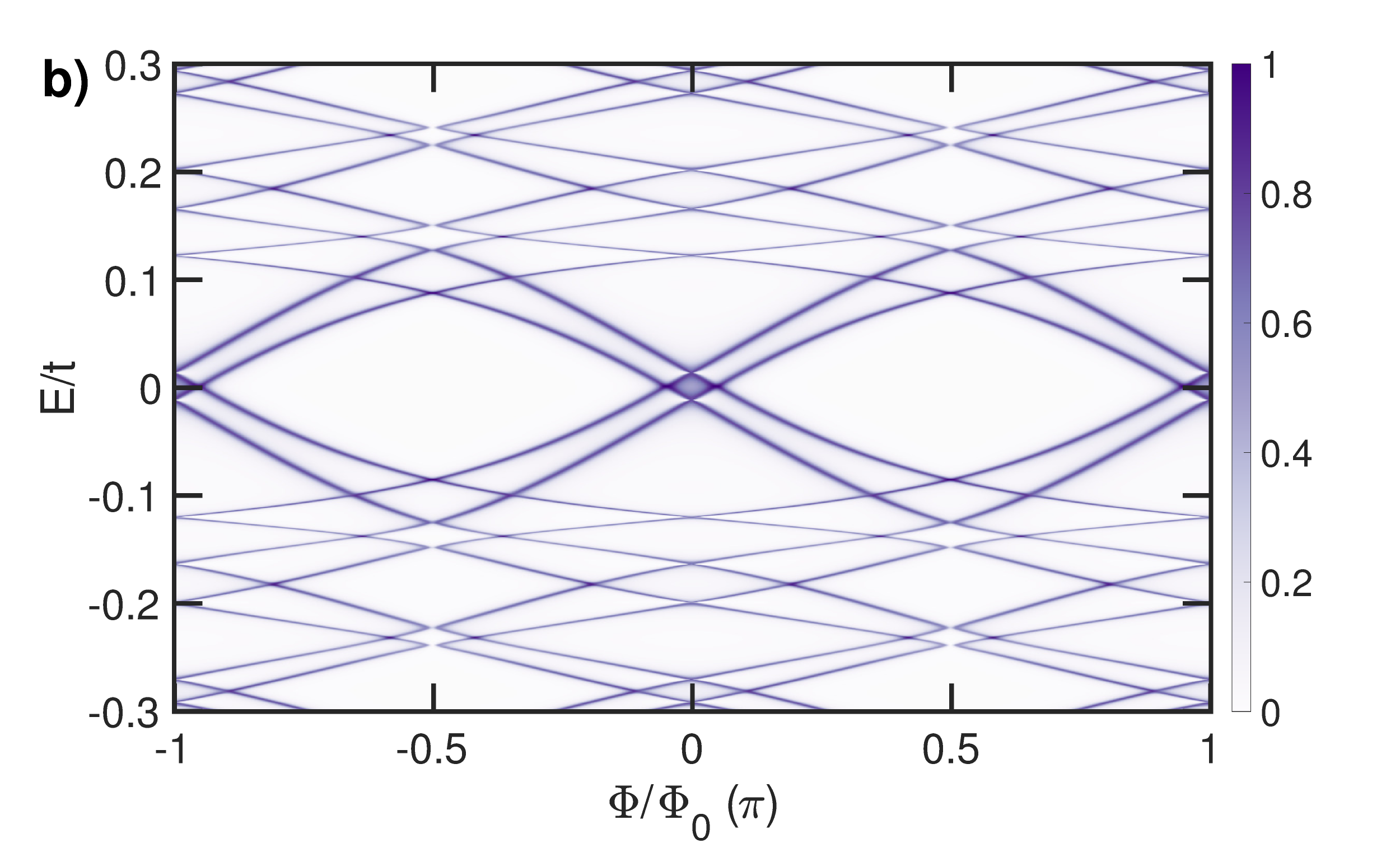}
		\includegraphics[width = \columnwidth]{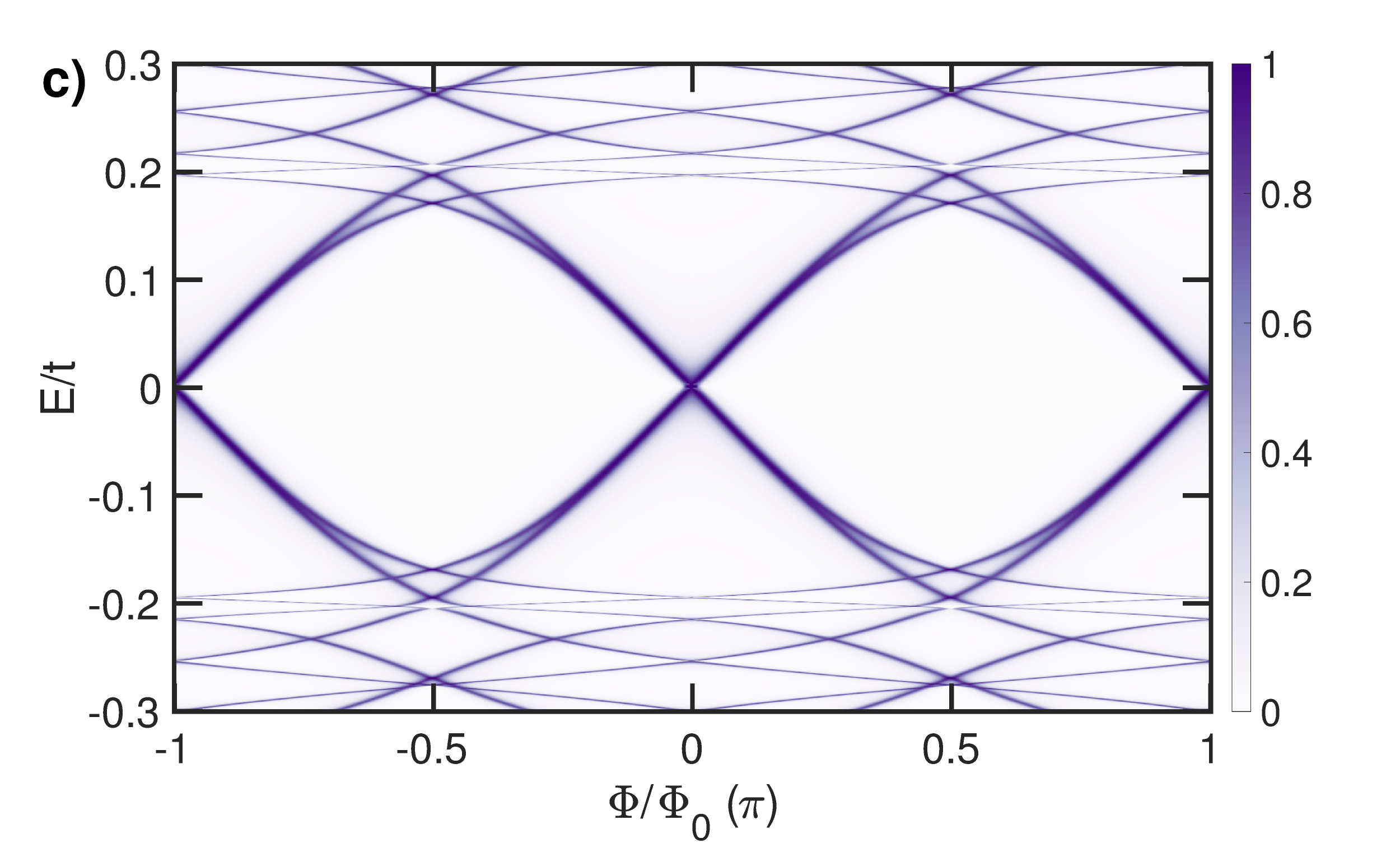}
		\caption{Transmission probability of an AB ring as a function magnetic flux $\Phi$ and energy $E$ for 
		various values of $p$-wave pairing amplitude (a) $\Delta/t = 0$, (b) $\Delta/t = 0.05$ and (c) $\Delta/t = 
		0.1$. The AB ring has an on-site potential $\mu/t = 1.0$ and normal leads are attached with a coupling 
		strength of $\tau/t = 0.3$.}
		\label{fig:AB_effect_rings}
	\end{figure}

	Resonances in the transport spectrum of the ring are caused by weak links between the device and the 
	leads~\cite{Doornenbal2015}.
	Applying a magnetic field to the AB ring in the normal regime ($\mu/t = 1$, $\Delta/t = 0$), as in 
	Fig.~\ref{fig:AB_effect_rings}a, we see there is completely destructive interference at $\Phi/\Phi_{0}= 
	\pm\frac{\pi}{2}$, which results in the transmission probability going to zero at these values of magnetic flux.
	The symmetry of the transmission resonances in Fig.~\ref{fig:AB_effect_rings} about zero energy follows 
	directly from the PHS.
	One important consideration is whether such a system would destroy superconductivity in the ring by applying a magnetic field larger than the critical field. 		
	To investigate this we can consider a ring of niobium titanium nitride as an example. If the ring has an area of 1~$\mu$m$^2$, one magnetic flux quanta corresponds to 2 mT, which is considerably smaller than a typical critical field of approximately 28 mT. This would allow the mapping of AB interference for many oscillation periods without collapsing the superconducting state\cite{Bosland1992,DiLeo1990}.
	
	In the case where $\Delta/t = 0.05$ and $\mu/t = 1$, as in Fig.~\ref{fig:AB_effect_rings}b, we see the 
	formation of a sub-gap state that oscillates in energy, as a function of magnetic flux, with a period of $\pi$.
	Note that the value of $\Delta/t$ is not large enough to overcome finite size effects here and that the 
	transport spectrum is characterised by a lack of the zero energy state at $\Phi/\Phi_{0} = 0$.
	Furthermore for a magnetic flux through the ring of $\Phi/\Phi_{0} = \pm\frac{\pi}{2}$ there still exists channels available for the conduction of electrons through the device which implies the interference is only partially destructive.
	This is in contrast to the case of the normal ring (shown in Fig.~\ref{fig:AB_effect_rings}a) where transmission is completely absent for this magnetic field strength.
	
	We now consider the transmission probability for an AB ring in the topologically non-trivial regime with an 
	applied magnetic field ($\Delta/t = 0.1$, $\mu/t = 1$) shown in Fig.~\ref{fig:AB_effect_rings}c.
	This is characterised by the presence of a zero energy state at $\Phi/\Phi_{0} = 0$.
	The MZMs exhibit a linear energy dependence at magnetic flux about $\Phi / \Phi_{0} = n\pi$ where $n \in 
	\mathbb{Z}$.
	Just as in Fig.~\ref{fig:AB_effect_rings}b, the ground state of the system is protected against destructive interference at $\Phi/\Phi_{0} = \frac{\pi}{2}$.
	However now this behaviour extends to the quasi-particle states at both higher and lower energies ($|E|/t \gtrsim 0.2$).
	Although the splitting of the zero energy state shows the degeneracy of the ground state is broken by an 
	applied magnetic field, these states are still protected against destructive interference, which is something 
	that is absent in the topologically trivial regime.
	
	\begin{figure}[b!]
		\includegraphics[width=\columnwidth]{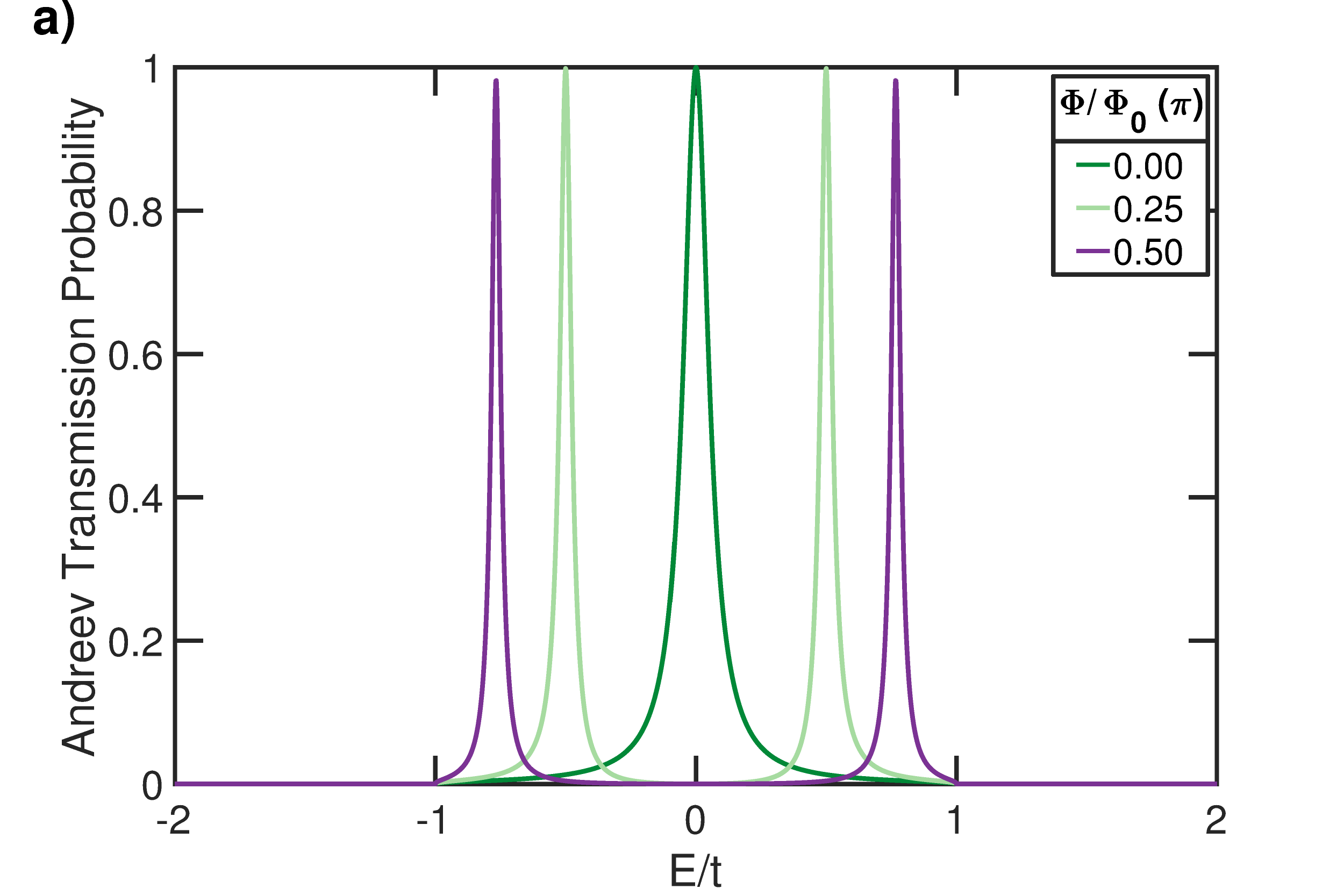}
		\includegraphics[width=\columnwidth]{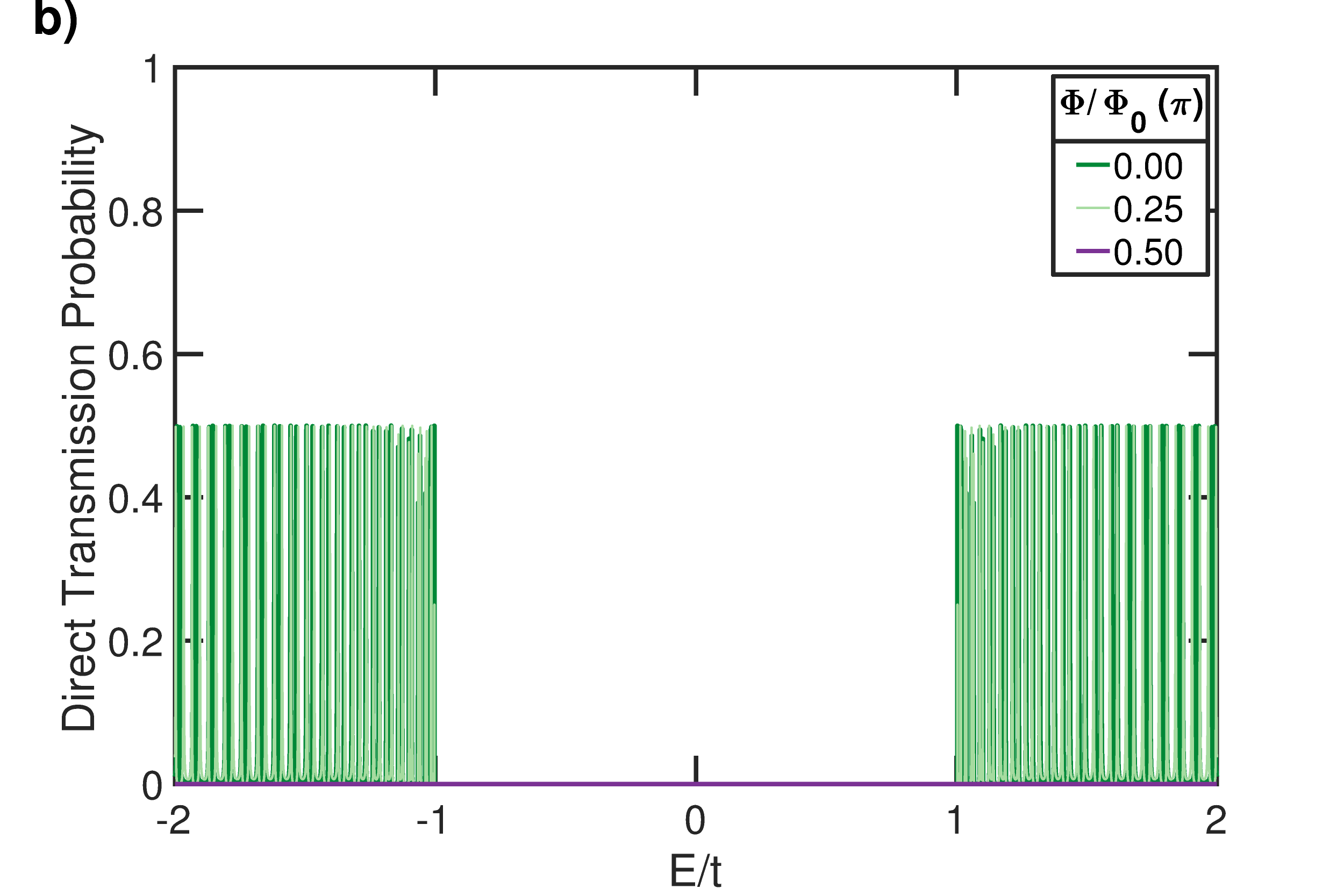}
		\caption{Contributions to the transport probability from (a) Andreev and (b) direct transmission for an AB 
		ring with $50\times2$ sites as a function of energy $E$. Here the on-site potential is $\mu/t = 1$. The AB 
		ring has a $p$-wave pairing amplitude $\Delta/t = 1.0$ and normal leads are attached with a coupling 
		strength of $\tau/t = 0.3$.}
		\label{fig:transmission_breakdown_2}
	\end{figure}
	
	\begin{figure}[b!]
		\includegraphics[clip,width=\columnwidth]{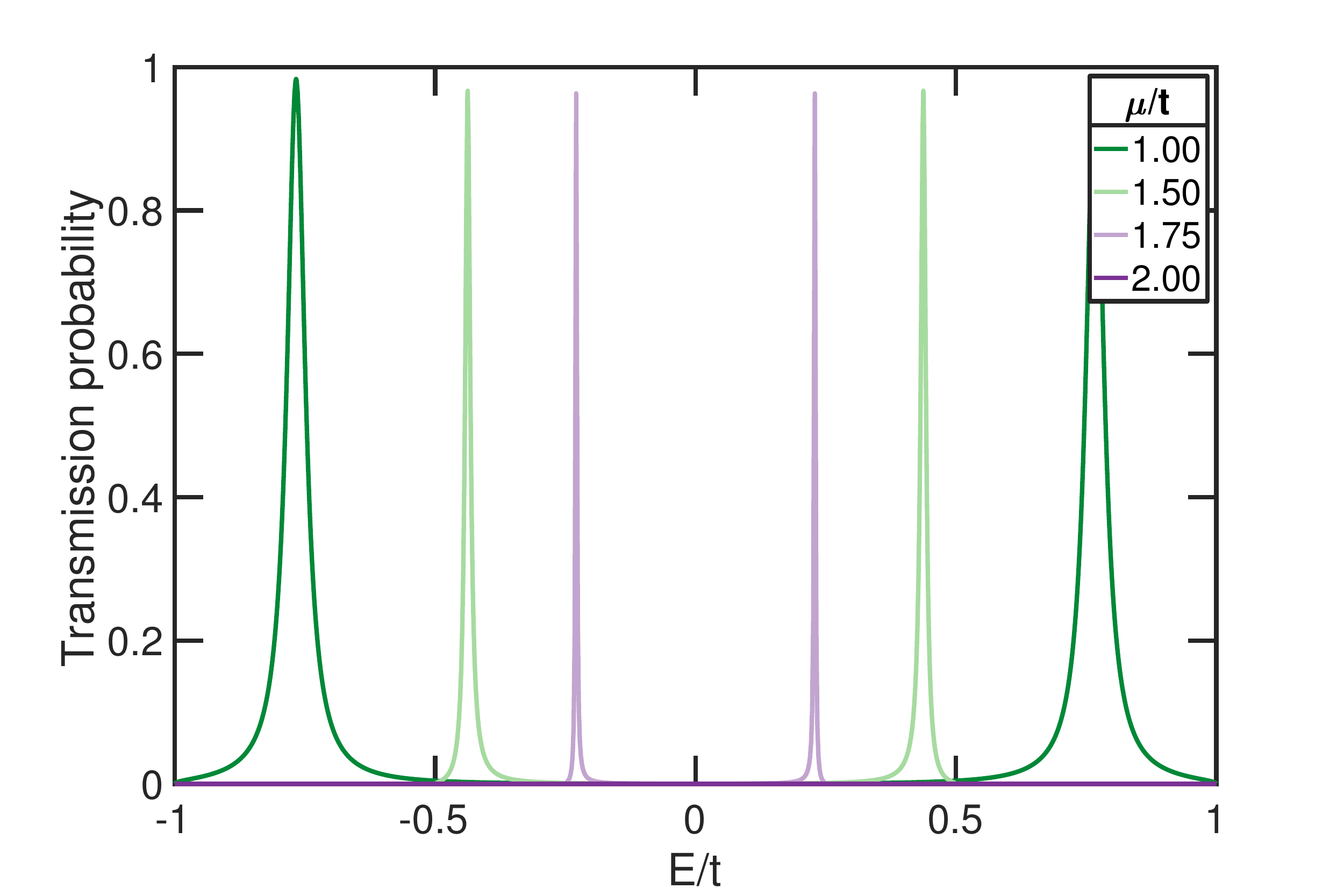}
		\caption{Transmission probability of an AB ring with $50\times2$ sites as a function of energy $E$ for 
		various on-site potentials $\mu$ and a magnetic flux of $\Phi/\Phi_{0}=\frac{\pi}{2}$. The AB ring has a 
		$p$-wave pairing amplitude $\Delta/t = 1.0$ and normal leads are attached with a coupling strength of 
		$\tau/t = 0.3$. When $\mu/t = 2$ the transmission is zero for all energies.}
		\label{fig:slices_of_delta}
	\end{figure}
	
	Fig.~\ref{fig:transmission_breakdown_2} gives the Andreev and direct transmission for an AB ring when 
	$\mu/t = 1$.
	Crossed Andreev transmission is absent from this figure because it is found to be zero for these parameters.
	At a magnetic flux $\Phi/\Phi_{0} = 0$ we see a zero energy peak for Andreev transmission 
	(Fig.~\ref{fig:transmission_breakdown_2}a), as well as contributions to the direction transmission from 
	quasi-particle states (Fig.~\ref{fig:transmission_breakdown_2}b).
	As the magnetic flux through the ring is varied such that the condition for completely destructive interference 
	is met ($\Phi/\Phi_{0} = \frac{\pi}{2}$) we see the suppression of all direct transmission, which is consistent 
	with the behaviour typically observed in AB interference.
	In contrast to this the Andreev transmission is found to persist at $\Phi/\Phi_{0} = \frac{\pi}{2}$.
	As with Fig.~\ref{fig:AB_effect_rings} we see a splitting in the zero energy state that depends on the magnetic 
	flux through the ring and increases with $\Phi/\Phi_{0}$ between 0 and $\frac{\pi}{2}$.
	This highlights it is the Andreev transmission that is protected against destructive interference in the AB ring.
	The extent to which Andreev transmission is protected against destructive interference is shown in 
	Fig.~\ref{fig:slices_of_delta}, where the transmission probability is plotted as a function of energy for several 
	different values of the on-site potential and a magnetic flux $\Phi/\Phi_{0} = \frac{\pi}{2}$.
	As the value of the on-site potential is increased and the system moves toward the topologically trivial 
	regime, the width of the sub-gap states becomes vanishingly small.
	
	The absence of destructive interference for an AB ring in the non-trivial regime, as shown in 
	Figs.~\ref{fig:AB_effect_rings}c~and~\ref{fig:slices_of_delta}, is of particular interest.
	Previous studies have concluded that electrons move through MZMs preserving phase 
	coherence~\cite{Fu2010}.
	This would suggest that transport should be suppressed at half integer multiples of magnetic flux quanta 
	(e.g. $\Phi/\Phi_{0} = \pm\frac{\pi}{2}$), as is typically true for the AB effect.
	Instead we see a splitting in energy of the topologically non-trivial state, and a shift in the energy of this state 
	away from zero, in response to changing the magnetic flux threading the loop.
	This raises questions as to whether the topological protection of the MZMs is still preserved in such a 
	situation.
	It also suggested the response of such a circuit to a magnetic field can be used as a probe of topologically 
	trivial Andreev bound states.
	
	\section{Effects of disorder}
	Experimentally realisable systems will contain noise which may perturb the system, one way this can occur is through disorder within the system itself. 
	To study the effect of disorder on our AB rings, we introduce to each site $n$ a quasi-random on-site energy $w_n$ which is obtained from a uniform distribution\cite{Gergs2016} centred about zero such that $w_n \in (-1, 1)$. The magnitude of the disorder is controlled via a disorder amplitude $P$ such that the on-site term for Eq.~\ref{eq:KitaevToyH} becomes $\mu_n = \mu + \left(P\cdot w_n\right)$.

	Electron transport in nanoscale devices is often limited by random offset charges stemming from surfaces and interfaces within the device\cite{Pourkabirian2014,Zorin1996,Zimmerman2008,Cedergren2017,Muller2019,Gustafsson2013}. This background charge disorder can either be static (approximately constant once the device is cooled to base temperature) or dynamic during the measurement period, resulting in an ensemble averaged signal over multiple disorder realisations.
	
	\begin{figure}[t!]
		\centering
		\includegraphics[width=1\columnwidth]{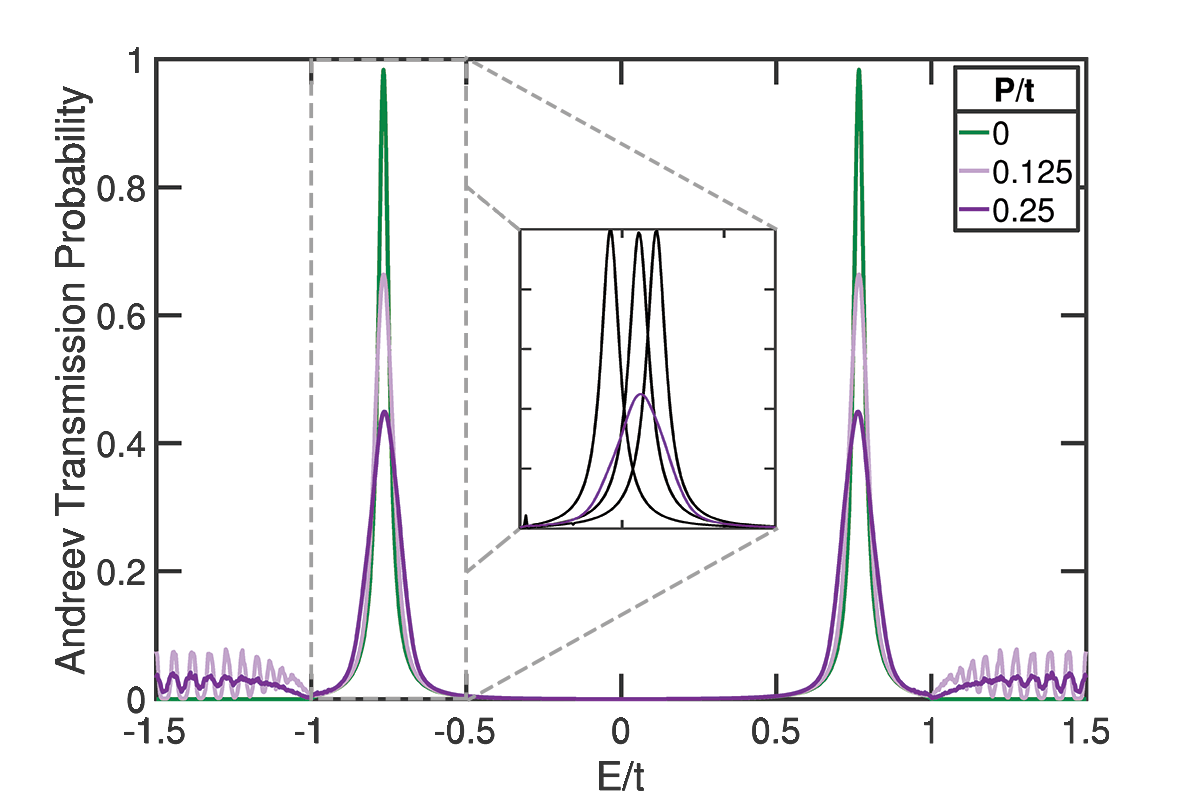}
		\caption{Transmission in an AB ring with $\Delta/t = \mu/t = 1.0$, 50 sites and a magnetic field of $\phi/\phi_0 = \frac{\pi}{2}$, for several values of disorder strength P. Each disorder value in this figure has been averaged over 1000 configurations. The inset shows the a sample of the static disorder realisations together with the dynamic disorder result for a disorder strength of $P/t = 0.25$.} \label{fig:disorder}
	\end{figure}
	
	For an AB ring without an applied magnetic field transmission is unchanged by the inclusion of weak disorder.
	The MZM in such a system is robust against both weak dynamic and static disorder, as seen in literature for 1D wires containing MZMs\cite{Lieu2018}.
	However the same is not true for an AB ring in the presence of a magnetic field.
	Fig.~\ref{fig:disorder} shows the effect of introducing a dynamic disorder scheme into the AB ring with magnetic field applied. The disorder has been introduced in the topologically non-trivial regime where $\phi/\phi_0 = \frac{\pi}{2}$ in the form of a random on-site variation.
	We see that increasing the strength of the disorder broadens the transmission peak and reduces the peak height of the system with dynamic disorder.
	This reduction in averaged peak height is caused by a disorder induced shifting of the energy values required to achieve peak transmission.
	This effect is illustrated in the inset, which shows the average transmission for a disorder strength of $P/t = 0.25$, with several of the static disorder realisations which have been averaged to get the dynamic disorder result.
	Note that the peak height for static disorder is unchanged by the strength of the disorder.

	The movement of the peaks in the inset of Fig.~\ref{fig:disorder} under disorder may be due to the fluctuations in the on-site potential which causes regions of the ring to transition either deeper into or out of the topologically non-trivial regime. 
	For a system being pushed deeper into the topologically non-trivial regime, the superconducting gap would become larger as in Fig.~\ref{fig:eigenspectra_halfpi_nontrivial}.
	For a system with magnetic field applied, we see in Fig.~\ref{fig:slices_of_delta} that decreasing the on-site potential causes the peaks to move to higher energies.
	However, that there still exists conductance channels with an applied magnetic field of $\phi/\phi_0 = \frac{\pi}{2}$ and in the presence of disorder suggests that these channels are robust against weak disorder.
	
	As the type of disorder affects the resultant transmission peaks, one would expect experimental signatures of the type seen in Fig.~\ref{fig:AB_effect_rings}c to change depending on the type of disorder introduced. 
	Static disorder would cause the separation between the transmission peaks to fluctuate, with the fluctuations increasing in magnitude with disorder strength.
	However for a system with dynamic disorder, one would expect the separation between the transmission peaks to remain relatively constant, but rather the peak amplitude would be decreased.
	In both cases, for weak disorder the conduction channels persist in the presence of partially destructive AB interference.
	
	\section{Conclusion}\label{sec:conclusion}
	
	We used the NEGF formalism to study the transport properties of an AB ring comprised of two Kitaev chains 
	coupled at either end by a normal link.
	We observed the effect of electron (hole) interference on the zero-bias transmission of this AB ring, which is 
	present when the magnetic flux through the ring is nonzero.
	We have shown how the MZMs in this AB ring change with on-site potential $\mu$, the $p$-wave pairing 
	amplitude $\Delta$, and the magnetic flux through the ring $\Phi$. 
	Each of these parameters has been shown to have a unique effect on the MZMs. 
	Having control of the physical parameters $\mu$, $\Delta$, and $\Phi$ in an experiment therefore allows the 
	MZMs to be probed in a controllable way.
	
	Furthermore, we have observed the transmission probability of electrons and holes through an AB ring with 
	MZMs to be persistent even when the condition for completely destructive interference are met.
	The power of the NEGF formalism as a tool for investigating this phenomenon is highlighted where the 
	transmission probabilities are divided into three separate contributions: Andreev and crossed Andreev 
	reflection and direct transmission.
	We have thereby demonstrated that transmission through MZMs in an AB ring is mediated by Andreev 
	reflection at the normal/superconductor interface. The dependence of the AB interference on the parameters 
	of the nanowires is therefore a useful probe of Majorana fermions in condensed matter systems.
	
	The mechanism which provides the above gap states protection from the AB remains unclear. We believe this issue requires further investigation using different tools to those used here and therefore we leave it for further work. 
		
	\begin{acknowledgements}
		
		The authors acknowledge useful discussions with S. Rachel, as well as MATLAB code provided by Jesse Vaitkus and Martin Cyster. 
		This work was supported by the Australian Research Council through grants CE170100039 (FLEET) and CE170100009 (EQUS). 
		CM is supported by the Swiss National Science Foundation through NCCR Quantum Science and Technology (QSIT). 
		BM would like to acknowledge funding from the Ministry of Human Resource Development (MHRD), Government of India, Grant no. STARS/APR2019/NS/226/FS under the STARS scheme and the Visvesvaraya Scheme of the Ministry of Electronics and Information Technology (MeiTy), Government of India. 
		This research was undertaken with the assistance of resources from the National Computational Infrastructure, which is supported by the Australian Government.
		
	\end{acknowledgements}
	
	\appendix
	
	\section{Particle-hole symmetry}\label{appendix:ParticleHoleSymmetry}
	
	The Kitaev Hamiltonian when expressed in the BdG form exhibits particle-hole symmetry (PHS).
	PHS represents a symmetry in the behaviour of electrons and holes and leads to a symmetry in the 
	eigenspectrum around zero energy.
	As we are investigating a superconducting system, the Hamiltonian contains a coupling term between 
	particles and holes, which is best described in the BdG formalism. 
	Employing the Nambu basis, we group creation and annihilation operators together as
	\begin{align}
	C^\dagger &= \{ c^\dagger_1 , \cdots , c^\dagger_n , c_1, \cdots ,c_n \}, \\
	C &= \{c_1, \cdots ,c_n , c^\dagger_1 , \cdots , c^\dagger_n\}^{T}.
	\end{align}
	
	This notation then allows us to write the BdG Hamiltonian as
	\begin{align}
	\hat{H} = \frac{1}{2} C^\dagger H_{\mathrm{BdG}} C
	\end{align}
	with the BdG Hamiltonian coefficient matrix for the Kitaev chain:
	\begin{align}
	H_{\mathrm{BdG}} =
	\begin{bmatrix}
	H & \Delta \\
	-\Delta^* & -{H^{*}}
	\end{bmatrix}.
	\end{align}
	
	Here PHS manifests directly in the symmetry between the particle and hole sectors of the Hamiltonian.
	
	\section{Derivation of the components of current in a system with particle-hole symmetry }\label{appendix:ComponentsOfTransmission}
	
	The conventional equation for calculating the transmission of electrons through a device is given in 
	Eq.~(\ref{eq:Caroli}).
	When applied to a system in conjunction with the BdG form of the Hamiltonian we must also take into 
	consideration the transmission probability for holes.
	This allows for the study of more transport modes than just the conventional transport of electrons through a 
	device~\cite{Sriram2019,Wei2001,Sun1999}.
	Andreev and crossed Andreev transmission allow for the reflection of an electron from an interface as a hole 
	both locally and non-locally respectively~\cite{Law2009,Wang2016,Nilsson2008}.
	In order to probe the role these processes play in this work, Eqs.~(\ref{eq:T_D})-(\ref{eq:T_CA}) were applied. 
	Although here we discuss how Eqs.~(\ref{eq:T_D})-(\ref{eq:T_CA}) can be derived from the Landauer current equations, it is important to note that they are themselves particle transmission functions and not current equations. 
	The distinction being that the transmission functions are independent of the Fermi distribution function (and therefore temperature and lead Fermi levels).
	These equations are derived from the Landauer current equation:
	\begin{align}
	I_\mathrm{L} = \frac{q}{h}\text{Tr}\left[-i\Sigma^{<}_\mathrm{L} A - \Gamma_\mathrm{L} G^{n}\right]
	\end{align}
	where $\Gamma_{\alpha} = \Gamma_{\alpha}(E)$ is the broadening matrix for lead $\alpha$. 
	We also have
	\begin{align}
	G^{n} = -iG^{<} = -iG^{R} \left(\Sigma^{<}_\mathrm{L} + \Sigma^{<}_\mathrm{R}\right) G^{A}.
	\end{align}
	and the spectral function is defined as
	\begin{align}
	A = G^{R} \Gamma G^{A}.
	\end{align}
	Applying these definitions to the Landauer current equation, we arrive at
	\begin{align}\label{eq:Landauer_inscatter}
	I_{\mathrm{L}} = \frac{q}{h}\text{Tr}[&-i\Sigma^{<}_\mathrm{L} G^{R} \left(\Gamma_{\mathrm{L}} + 
	\Gamma_{\mathrm{R}}\right) G^{A}\nonumber \\
	&+ i\Gamma_{\mathrm{L}} G^{R} \left(\Sigma^{<}_\mathrm{L} + \Sigma^{<}_\mathrm{R}\right) G^{A}].
	\end{align}
	As we are working in the BdG formalism which utilizes the Nambu (particle-hole) basis, we must take both 
	particle and hole transmission into account.
	With the application of normal leads there is no coupling between particles and holes and as such the 
	in-scattering matrix ($\Sigma^{<}$) becomes block diagonal (one block for electron scattering and one for 
	hole scattering).
	We are therefore able to write the in-scattering matrix for lead $\alpha$ as
	\begin{align}\label{eq:inscatter}
	\Sigma^{<}_{\alpha} = i\Gamma^{e}_{\alpha}f^{e}_{\alpha} + i\Gamma^{h}_{\alpha}f^{h}_{\alpha}
	\end{align}
	where the superscript e(h) denotes the electron (hole) component in the Nambu basis and ${f_{\alpha}(E)=(e^{(E+E_{F})/kT}+1)^{-1}}$ is the Fermi distribution for the $\alpha$-lead with Fermi energy $E_{F}$ at temperature $T$.
	Combining Eq.~(\ref{eq:Landauer_inscatter}) and Eq.~(\ref{eq:inscatter}) we arrive at
	\begin{align}
	I_{\mathrm{L}}^{e(h)} = \frac{q}{h}\text{Tr}&[(\Gamma^{e(h)}_{\mathrm{L}}f^{e(h)}_{\mathrm{L}} + 
	\Gamma^{h(e)}_{\mathrm{L}}f^{h(e)}_\mathrm{L}) G^{R} \left(\Gamma_\mathrm{L} + 
	\Gamma_\mathrm{R}\right) G^{A}\nonumber \\
	-&\Gamma_{\mathrm{L}} G^{R}((\Gamma^{e(h)}_{\mathrm{L}}f^{e(h)}_{\mathrm{L}} + 
	\Gamma^{h(e)}_{\mathrm{L}}f^{h(e)}_{\mathrm{L}})\nonumber \\
	+&(\Gamma^{e(h)}_{\mathrm{R}}f^{e(h)}_{\mathrm{R}} + 
	\Gamma^{h(e)}_{\mathrm{R}}f^{h(e)}_{\mathrm{R}})) G^{A}].
	\end{align}
	which after some algebraic manipulation returns
	\begin{align}
	I^{e(h)}_{\mathrm{L}} = 
	\frac{q}{h}\text{Tr}\bigg[&\Gamma_{\mathrm{L}}^{e(h)}G^{R}\Gamma_{R}^{e(h)}G^{A}\left(f_{\mathrm{L}}^{e(h)}
	 - f_{\mathrm{R}}^{e(h)}\right)\nonumber \\
	+ &\Gamma_{\mathrm{L}}^{e(h)}G^{R}\Gamma_{\mathrm{L}}^{h(e)}G^{A}\left(f_{\mathrm{L}}^{e(h)} - 
	f_{\mathrm{L}}^{h(e)}\right)\nonumber \\
	+ &\Gamma_{\mathrm{L}}^{e(h)}G^{R}\Gamma_{\mathrm{R}}^{h(e)}G^{A}\left(f_{\mathrm{L}}^{e(h)} - 
	f_{\mathrm{R}}^{h(e)}\right)\bigg].
	\end{align}
	As we only want to study the transmission, rather than the current we can neglect the constant factor as well 
	as the Fermi windowing function, after which we arrive at the transmission equations given in 
	Eqs.~(\ref{eq:T_D})-(\ref{eq:T_CA}).
	
	\bibliography{./main}
	
\end{document}